\documentclass{amsart}

\newtheorem{problem}{P\#}

\usepackage{graphicx}
\newcommand{\nest}[1]{\left(#1\right)}
\newcommand{\f}[2]{#1\nest{#2}}

\begin{document}
\title{Remarks on d-Dimensional TSP Optimal Tour Length Behaviour}
\author{Andrei G. Yaneff}
\address{E.T.Consulting, 6 Buntovnik street, Sofia, 1421, Bulgaria}
\email{agy@techno-link.com}
\thanks{E.T.C.2/2002}
\date{August 14, 2002 and, in revised form, September 18, 2002}



\begin{abstract}
The well-known $O\nest{n^{1-1/d}}$ behaviour of the
optimal tour length for TSP in d-dimensional Cartesian space
causes breaches of the triangle inequality. 
Other practical inadequacies of this model are discussed,
including its use as basis for approximation of the TSP optimal
tour length or bounds derivations, which I attempt to remedy.
\end{abstract}

\maketitle
\section{Introduction}
Studies of the behaviour of the optimal tour value in respect of
the TSP has been in place since the fifties.
Considerate work has been done with the $O\nest{n^{1-1/d}}$
model of behaviour based on results of Beardwood, Halton and Hammersley
\cite{Beardwood}. The approach taken here is one of simple real analysis
and some of the intermediate results correspond with those of Snyder and
Steele \cite{142745} and Bern and Eppstein \cite{161018}.
\par
As pointed out by Bern and Eppstein \cite{161018}, an a priori bound
on a geometric graph is a bound that depends on the assumption
that all vertices lie within a given container. Rather than bounds,
I consider the behaviour of the optimal tour length of a geometric
graph in a 2D rectangular region, which is transformed to a square
and then simplified (extended) to d-dimensional cube. This is compared
to the behaviour of the MST length in the same container space.
\par
The problem of studying the behaviour or approximation of the optimal
tour length of TSP instances ($\|O\|$) can be illustrated by an example:
Given two TSP instances with substantially
different number of vertices have the same optimal tour length.
\par
It is shown that the general behaviour of $\|O\|$ does not depend on the
number of vertices. I also give experimental evidence in support of
this using a small set of TSPLIB \cite{TSPLIB} instances, and conclude
that the commonly cited $O\nest{n^{1-1/d}}$ behaviour is
inadequate in both description of behaviour (nor useful bounds) of
$\|O\|$. A simple alternative is suggested.
\par
The present is an extended version of a mention in \cite{Yaneff1}.
\par
\section{Terms and Definitions}
\par
The Minimum TSP in Cartesian space is a special case of the Minimum
Weight Hamiltonian Cycle problem. Both are specified as follows:
\begin{problem}[Minimum Hamiltonian Cycle\label{MinHC}]
Given  $G=(V,E)$ with a cost function $d:E \rightarrow R$,
find solution $O=(V_O,E_O) \subseteq G$
such that
\begin{enumerate}
\item[(a)] $O$ is a Hamiltonian cycle;
\item[(b)] $\|O\| = \sum_{e:E_O} d(e)$ is minimum;
\end{enumerate}
\end{problem}
\begin{problem}[Minimum TSP in Cartesian Space\label{MinCTSP}]
P\#\ref{MinHC} and
\begin{enumerate}  
\item[(a)] $\forall v:V \cdot \f{coords}{v}\in R^d$ and
\item[(b)] $\forall (u,v):E \cdot \f{d}{u,v} =
|\f{coords}{u}-\f{coords}{v}|$
\end{enumerate} where $coords: V \rightarrow R^d$ are Cartesian.
\end{problem}
\par
I (unnecessarilly) split the problem specification this way for my
personal convenience in what follows (separating construction and
geometric aspects of the problems), and for the following
observations (from P\#\ref{MinHC}):
\begin{enumerate}
\item[(a)]  $|V|=|V_O|=|E_O|$;
\item[(b)]  Solution bounds:
\begin{equation}
n w_0 \leq \|O\| \leq n w_1 \label{bounds}
\end{equation}
where $n = |E_O|$,\\
$w_0=\min \lbrace d(e)|e \in E\rbrace \land
w_1=\max \lbrace d(e)|e \in E\rbrace$
\end{enumerate}
\par
The notation used is fairly standard. I have attempted to maintain consistent
naming, e.g. $Oabc_i$ is the value (approximate) of the exact $\|O\|$, and the
subscript $i$ is the number of 'alternative' formulation of $Oabc$,
e.g. $Ob_1$ is an alternative to $Ob_2$.

\section{Two Similar Formulations of d-Dimensional $\|O\|$ Behaviour}
\par
Consider a rectangle (being a container space of the TSP graph and
respective tour) and that we have a uniform grid tight within (and on)
that rectangle with size $a$ by $b$ points, assuming without loss that
$a$ is even. The optimal tour value and the diagonal of the
rectangle are:
\begin{eqnarray}
\|O\|=a \cdot b \cdot w_0=n . w_0 \label{Oe0}\\
w_1 = w_0 \sqrt{(a-1)^2 + (b-1)^2} \label{w1a}
\end{eqnarray}
Dividing:
\begin{equation}
\frac{\|O\|}{w_1}=\frac{a \cdot b}{\sqrt{(a-1)^2 + (b-1)^2}} \label{Oe1}
\end{equation} 
\par
If $a=b$ than we have a square, and
$\|O\|/w_1=n/\nest{\nest{\sqrt n - 1}\sqrt 2}$.
To carry this to Cartesian d-dimensions, (\ref{Oe0}) and (\ref{w1a})
are changed to
$O_1 = w_0\prod_{i=1}^{d} a_i = n \cdot w_0 $ and
$w_1 = w_0 \nest{\sum_{i=1}^{d}(a_i-1)^2}^{1/2}$.
Assuming $ \forall i,j:1\dots d \cdot a_i=a_j$ then 
\[ 
\frac{O_1}{w_1} = \frac{n}{(n^{1/d}-1)\sqrt{d}} 
\]
\par
Note that $O_1$ differs from the commonly cited $O(n^{1-1/d})$
optimal tour length behaviour, which can be formulated by taking 
$w_1 = w_0\nest{\sum_{i:1\dots d}{a_i}^2}^{1/2} $ (under the assumption
that $a$ is large), then:
\[ \frac{O_2}{w_1} =\frac{n^{1 - 1/d}}{\sqrt{d}} \]
\par
It should be noted that in the unit d-dimensional cube $w_1=\sqrt d$.
\par
For a metric (non-topological) graph $\|O\| \geq 2 w_1$ must hold
\cite{Reingold}. Both $O_1$ and $O_2$ fail
to preserve the triangle inequality when $d=1$, due to the "spatial"
nature of the derivations (no container). It is also evident that
$O_1$ preserves the triangle inequality for any $n\geq 2$,
whereas $O_2$ fails in doing so when $n^{1-1/d} \leq 2 \sqrt d$.
(There is nothing to prevent such instances from occuring.)
The ratio of the two estimates as
$d \rightarrow \infty$ is:
\begin{equation}
\f{e}{n,d} = \frac{O_2}{O_1}
= 1-\frac{1}{n^{1/d}} \rightarrow 0 \label{e}
\end{equation}
\par
\begin{figure}
\includegraphics[scale=0.7]{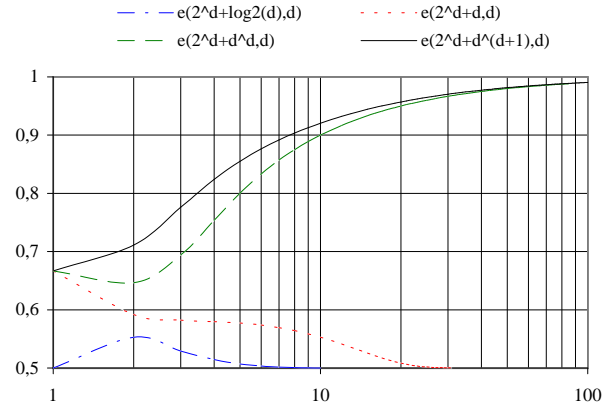}
\caption{Plots of $\f{e}{2^d+k,d}$, for $k=\log_2 d$,
$k=d^{\log_2 d}$, $k=d^d$, and $k=d^{d+1}$, for $d = 1 \dots 100$.}
\label{fig1}
\end{figure}

The divergent behaviour of $O_1$ and
$O_2$ is fixed by imposing the
condition $n\geq2^d + k$, $k \geq 0$ in order to have a valid d-dimensional cube,
and $ 1/2 \leq \f{e}{n,d} \leq 1$.
Either $O_1$ or $O_2$ will fail in the description (if they are
assumed to be descriptive) of the optimal tour length behaviour (see Fig.\ref{fig1}).
\par
The MST length behaviour ($\|T\|$) is analogous 
\cite{98596} to that of the TSP in a grid:
$\|T\| = \left(n-1\right) \cdot w_0$ and
$T_1/w_1 = \left(n-1\right)/\left((n^{1/d}-1)\sqrt d\right)$.
$\|O\|$ can be expressed in terms of $\|T\|$ (treated as known
value computed exactly by Prim's algorithm, say):
\[
Ot_1 = \|T\| + \frac{w_1}{\nest{n^{1/d}-1}\sqrt d}
\]
\[
Ot_2 = \|T\| + \frac{w_1}{n^{1/d}\sqrt d } 
\]
\par
As a matter of fact:
\begin{eqnarray}
\|O\| \geq \|T\| + w_0 \label{lb2}\\
\nest{2\|T\| \geq} \|O\| \geq \|T\|\frac{n}{n-1} \label{lb3}
\end{eqnarray}
\par
It is the case that for any metric graph
$\|T\|$ accounts for at least 50\% of $\|O\|$ and can be
considered a major term in any attempted approximation.
It is obvious that the second term in $Ot_1$ may well lead to
$w_0 > w_1/\nest{\nest{n^{1/d}-1}\sqrt d}$, as
$n \rightarrow \infty$, and ultimately $Ot_1 \rightarrow \|T\|$ (so will $Ot_1$),
which cannot be. This suggests that $\|O\|$ does not depend on
the number of vertices. Moreover, the $\f{e}{n,d}$ term also
remains in the relative difference in estimates for the remaining 50\% of
$\|O\|$, as seen from $Ot_1$ and $Ot_2$. In other words, if
$\|O\|$ were to depend on the number of vertices the following must be true:
\[
\f{e}{n,d} = \frac{Ot_2-\|T\|}{Ot_1-\|T\|}\
= \frac{O_2}{O_1} \land \|T\| \neq 0\]
\par
\begin{figure}
\includegraphics[scale=0.7]{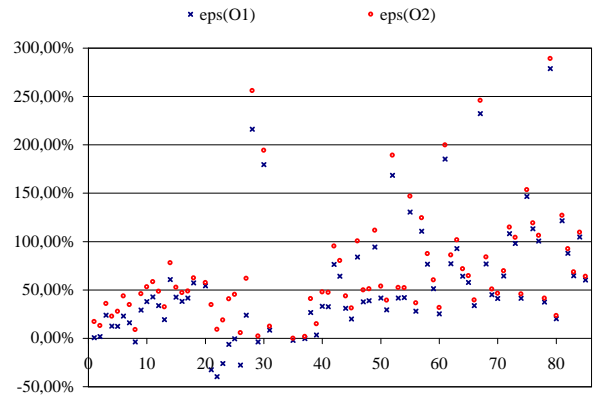}
\caption{Point histograms of relative error of $O_1$ and
$O_2$ - 85 instances}
\label{fig2}
\end{figure}
\begin{figure}
\includegraphics[scale=0.7]{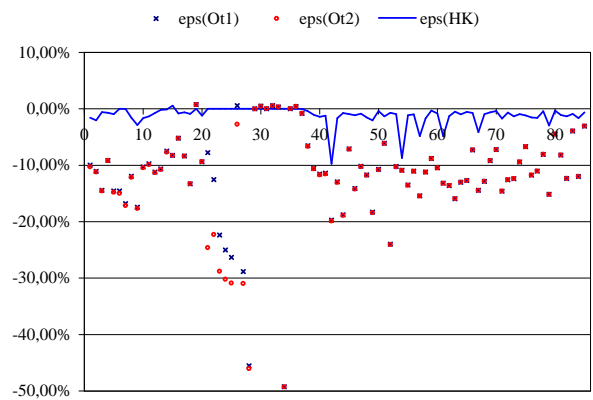}
\caption{Point histograms of relative error of $Ot_1$, $Ot_2$
 and $HK$ bound (-line used for clarity) }
\label{fig3}
\end{figure}

Obviously this is absurd, unless $Ot_1=Ot_2\land O_1=O_2$, which is not
the case. Thus, it is difficult to contemplate using $O_1$ or $O_2$
(see Fig.\ref{fig2}) as basis for behaviour description of
$\|O\|$. The $Ot_i$'s are slightly better (see Fig.\ref{fig3}), exhibiting
an almost "bounding" behaviour, for which we already have (\ref{lb2}) and
(\ref{lb3}) and the $HK$ bound (see Fig.\ref{fig5}).
\par
\begin{figure}
\includegraphics[scale=0.7]{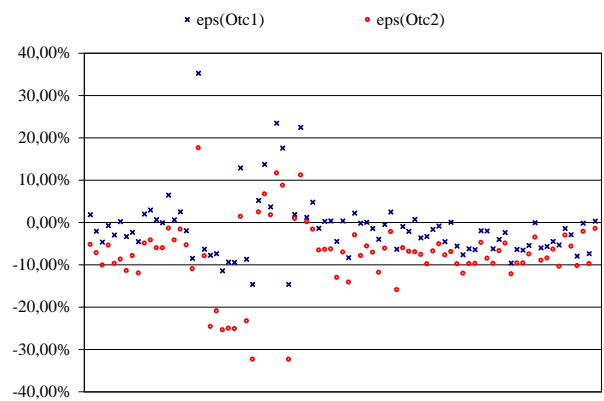}
\caption{Point histograms of relative error of $Otc_1$ and $Otc_2$}
\label{fig4}
\end{figure}
\begin{figure}
\includegraphics[scale=0.7]{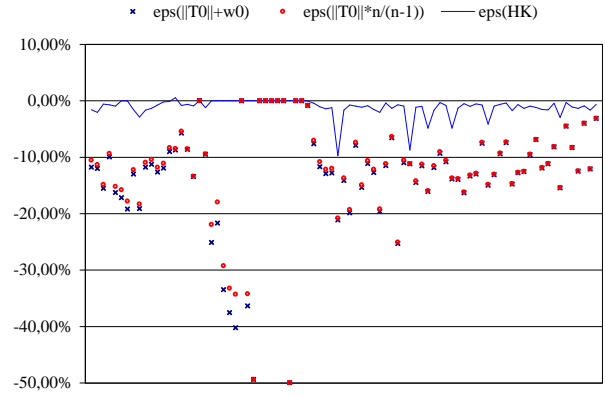}
\caption{Point histograms of the lower bounds $\|T\|+w_0$,
 $\|T\|\frac{n}{n-1}$ and  $HK$ bound (-line used for clarity)}
\label{fig5}
\end{figure}

From the above one can conclude that $n$ can do more harm than good in the
analysis of $\|O\|$. Setting $n=2^d$ (see above), yet another
approximation (see Fig.\ref{fig4} and Table.\ref{ostats}) to $\|O\|$ is obtained:
\[
Otc_1 = \|T\| + \frac{w_1}{\sqrt d} \nest{ = Otc_2 + \frac{w_1}{2\sqrt d}}
\]
\begin{table}[ht]
\renewcommand\arraystretch{1.5} 
\noindent\[
\begin{array}{|c|r r r r|}
\hline
X&\min \epsilon &\max \epsilon&\max \epsilon -\min \epsilon&\epsilon_{rms} \\
\hline
O_1&0&1668.14&1668.14&308.96\\
O_2&-39.69&1632.05&1671.74&299.91\\
\hline
Ot_1&-49.26&0.76&50.02&14.16 \\
Ot_2&-49.28&0.75&50.02&15.00 \\
\hline
Otc_1&-14.64&35.27&49.91&7.75 \\
Otc_2&-32.32&17.61&49.93&11.14 \\
\hline
HK^{*}&-9.73&0.55&10.28&2.18\\
\|T\|+w_0&-49.98&0&49.98&16.25\\
\|T\|\frac{n}{n-1}&-49.98&0&49.98&15.48\\
\hline
\end{array}
\]
\caption{Relative error ranges for the various formulas.
$\epsilon = 100\cdot \nest{X/\|O\|-1}\%$. (The largest $HK^*$ value
+0.55 is possibly due to a misprint in \cite{Helsgaun})}\label{ostats}
\end{table}

\subsection{Test Data}
\par
The test data consist of 85 instances in total of sizes $4\leq n \leq 2560$,
I have used 17 instances of my own and 68 instances from TSPLIB \cite{TSPLIB}
of type EUC\_2D, CEIL and ATT. The three norms were treated as
EUC\_2D without rounding. ATT are treated in the same way,
as tour values were multiplied by  $\sqrt{10}$ to bring them in line
with unrounded EUC\_2D. The HK-bound values are taken from \cite{Helsgaun}.
There were 5 TSPLIB with published shortest tour, for which I have been
able to find shorter tours. These are given in Table.\ref{shortertours}.
The HK bound is greater than the actual tour length for $tsp225$ and
$att48$ by $0.55\%$ and $0.01\%$ (see Table.\ref{shortertours}), respectively, possibly due to a
typographical error (or the use of initial rounding in computing the
HK relaxation). I do not consider accuracy (see Table.\ref{instances})
in this analysis of major
importance, as for the 50 instances for which the optimal tour itself
is not published, the data can be fixed (by point displacement) so
the reported values are accurate (This is not the case with HK bound).
\begin{table}[ht]
\renewcommand\arraystretch{1.5} 
\noindent\[
\begin{array}{|c|r r r r|}
\hline
Instances&Size&O&\|O\|&HK \\
\hline
17&4\leq n \leq 2560&known&exact&-\\
18&48\leq n \leq 2392&published&exact&reported\\
50&99\leq n \leq 2319&reported&reported&reported\\
\hline
\end{array}
\]
\caption{TSP instances}\label{instances}
\end{table} 
\begin{table}[ht]
\renewcommand\arraystretch{1.5} 
\noindent\[
\begin{array}{|c| r r r r|}
\hline
Instances&\|O_{reported}\|&HK&\|O_{published}\|&\|O_{shorter}\|\\
\hline
eil51&426&422.4&429.983307&429.117879\\
st70&675&670.9&678.597412&677.194473\\
eil101&629&627.3&642.309509&640.42177\\
ch130&6110&6074.6&6110.859375&6110.722217\\
ch150&6528&6486.6&6532.282227&6530.906184\\
\hline
\hline
att48&10628&10602.1&10601.1282&=\\
tsp225&3919&3880.3&3858.999064&=\\
\hline
\end{array}
\]
\caption{Shorter tours. Values are as reported in \cite{Helsgaun}
and computed from published tours in \cite{TSPLIB}}
\label{shortertours}
\end{table}
\section{Conclusions}
\par
Inadequacies of $O\nest{n^{1-1/d}}$ and similar formulations are
demonstrated. A probabilistic or statistical \cite{314081} argument
may show otherwise.
\par
The optimal tour length does not depend on the number of vertices - 
consider again a grid example in a square:
Let's suppose that we have found an optimal tour $S=(V_S,E_S)$ from $G=(V,E)$
(a comb-like shape, say) and insert (in $G$) exactly one zillion
equidistant points along each arc in $E_S \subseteq E$ of the optimal
tour $S$ to get $G_Z$. If $S_Z$ is an optimal tour of $G_Z$ then 
$\|S\|=\|S_Z\|$.
Such point insertion is not inconsequential - the number of optimal
solutions may be reduced, i.e. $G_Z$ has a lesser number of
optimal solutions than $G$.
\par
I make no claims as to quality (if used as estimates) of the above
given formulae, as I would not use any of them (see Table.\ref{ostats})
as an estimate of $\|O\|$. $O_1$ is an upper (easily proven so) bound,
for any $n\geq2$, however running the Nearest town algorithm \cite{Reingold} to get us an upper bound
within a factor of two is simpler. A simple derandomisation of the
same algorithm yields an approximate solution within a factor of
$\sqrt 2$ (provable under a geometric argument) for P\#\ref{MinCTSP}
in the plane.
\bibliographystyle{amsplain}
\bibliography{biblio}
\end{document}